# An efficient solution for accelerating very high intensity beams in the low and medium energy regime

Chuan Zhang[*, 1], Holger Podlech[2], Ulrich Ratzinger[2], Rudolf Tiede[2]

[1] *GSI Helmholtz Center for Heavy Ion Research, Planckstr. 1, Darmstadt, Germany*

[2] *Institute for Applied Physics, Goethe-University, Frankfurt a. M., Germany*

*Abstract*:

Taking advantage of the 0° synchronous phase, the KONUS ("Combined 0° Structure" translated from German "Kombinierte Null Grad Struktur") beam dynamics strategy enables long accelerating sections with lens-free slim drift tubes in the low and medium energy regime. It has successfully realized worldwide many room temperature H-type linacs with compact layouts and good beam performance. In this paper, a further development of this solution i.e. to combine the KONUS dynamics with the young superconducting CH (Crossbar H-Type) structure for accelerating very high intensity beams is being presented. The efficiency of the new solution has been shown by the systematic design studies performed for a 150mA, 6MW deuteron linac.

---

[#] e-mail: c.zhang@gsi.de

# I. INTRODUCTION

For the production of various useful high intensity secondary beams e.g. neutrons, the development of MW-class linear accelerators has become very attractive since several decades. Serving as drivers of large-scale facilities for modern scientific and civil applications, this kind of linacs usually need to deliver very powerful light ion beams e.g. $H^+$, $H^-$, $D^+$ to bombard a certain target. Having been put into operation already in the early 1970s, the LANSCE linac [1] formerly known as LAMPF can provide protons with an average beam power up to ~0.8MW. Different from the full room temperature (RT) LANSCE linac, the ~1.4MW Spallation Neutron Source (SNS) linac [2] built in 2006 started to employ the superconducting (SC) radio frequency (RF) technology for the beam acceleration in the high $\beta$ region. So far, many modern facilities based on this kind of high power driver linacs (HPDL) have been realized e.g. J-PARC [3] or proposed e.g. MYRRHA [4] worldwide, with the tendency to start the cold part already in the low and medium $\beta$ region.

The average beam power is given by Eq. (1):

$$\text{Average beam power} = \text{beam energy} \times \text{beam current} \times \text{beam duty factor} \quad (1)$$

To reach the beam power in the order of MW, there are typically the following several ways to combine these three factors for a modern HPDL machine:

- high energy $\times$ modest current $\times$ low duty factor: e.g. LANSCE and SNS.
- high energy $\times$ low current $\times$ continuous wave (CW): e.g. MYRRHA and PIP-II [5].
- low energy $\times$ very high current $\times$ CW: e.g. LEDA [6] and this study.

FIG. 1 shows a schematic layout of a modern large-scale HPDL. It can be roughly divided into the following three parts:

- Very low $\beta$ (0.01 ~ 0.1) part: the RT Radio-Frequency Quadrupole (RFQ) accelerator is a standard injector structure.

- High $\beta$ (>0.5) part: the SC elliptical cavity is dominating in this area.
- Low and medium $\beta$ (0.1 ~ 0.5) part: different solutions based on various RT and SC Drift-Tube Linac (DTL) structures can be used.

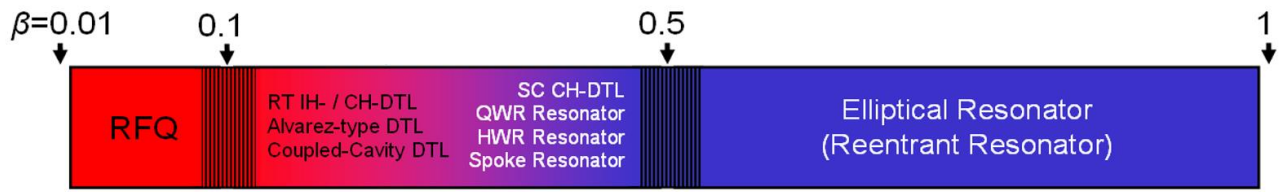

FIG. 1. Typical layout of a modern large-scale HPDL.

For proton and ion linacs, the classic beam dynamics strategy applies negative synchronous phases typically -30 ° ~ -40 ° to the accelerating cells. It provides the beam longitudinal stability but at the same time RF defocusing effects in the transverse planes. In case of high currents, space charge effects are also a big concern for the low and medium energy beams, especially when $\beta \leq 0.2$.

Based on the classic beam dynamics strategy, two well established solutions for the conventional low and medium energy linacs are as follows:

- Using long (multi-cell, here after referred to as >3 cells) RT structures e.g. Alvarez DTL with integrated magnetic lenses inside drift tubes.
- Using short (typically 2 ~ 3 cells per cavity) SC structures e.g. Half Wave Resonator (HWR) or Quarter Wave Resonator (QWR) with independent lenses outside of the cavities.

For both of them, relatively high magnet density is required to provide the accelerated beam sufficient transverse focusing.

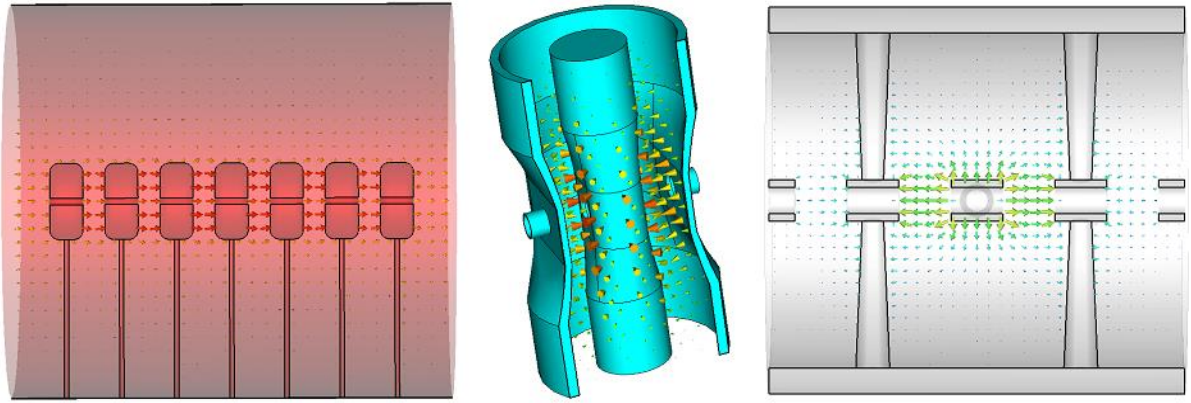

FIG. 2. Alvarez (left), HWR (middle) and H-type (right) structures with electric fields.

Different from the above-mentioned two solutions, a special beam dynamics strategy so-called KONUS (Combined 0° Structure) [7, 8] applies the 0° synchronous phase instead of negative synchronous phases to most accelerating cells. The purpose is to overcome the transverse defocusing effects by maximizing the acceleration efficiency so that the use of long accelerating sections with lens-free slim drift tubes becomes feasible.

However, the separatrix that exists at negative synchronous phases e.g. -30° will shrinks to zero at $\varphi_s = 0°$ (see FIG. 3). To solve this problem, KONUS uses only the area marked by blue arrows in the longitudinal phase space, which means that the synchronous particle defining the geometrical layout of the drift tube array and the bunch center (BC) particle are decoupled. The energy shift at the beginning of a 0° section is realized by setting $W_s < W_{BC}$, while the phase shift is obtained by adjusting the tank RF phase or the geometrical length of the transition cell when the transition happens between cavities or inside a cavity, respectively. In short, the beam is injected into a 0° section asynchronously with a surplus in bunch energy and with a proper phase slip against the synchronous particle.

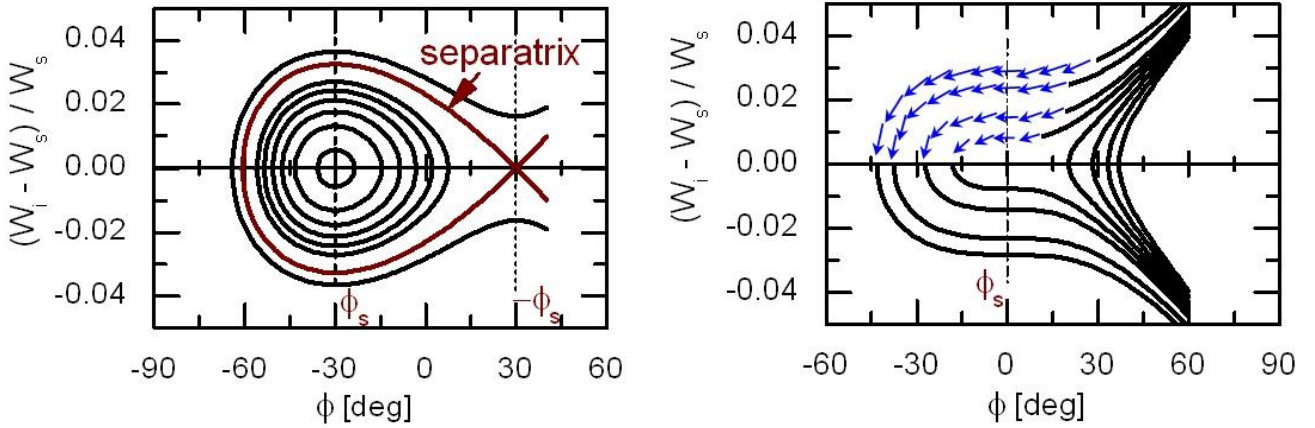

FIG. 3. Particle trajectories in the longitudinal phase space at $\varphi_s = -30°$ (left) or $\varphi_s = 0°$ (right).

This concept has been first applied in combination with an IH (Interdigital H-Type) structure for the heavy ion post-accelerator at the Munich tandem laboratory [9]. TABLE I shows a list of existing and planned KONUS-based H-Type DTLs worldwide. It can be seen that the development of this kind of machines is not only towards higher $\beta$ but also towards higher current. So far, all realized and to-be-realized KONUS machines are normal conducting accelerators.

TABLE I. An overview of KONUS-based H-Type DTLs.

| **Linac** | **Resonator** | **Ion** | **$f$ [MHz]** | **$I$ [mA]** | **$\beta$** | **Status** |
|---|---|---|---|---|---|---|
| GSI HLI | IH | $A/q \leq 9.5$ | 108.408 | 0.1 | 0.025 ~ 0.055 | in operation |
| GSI HSI | IH | $A/q \leq 65$ | 36.136 | $\leq 0.25*(A/q)$ | 0.016 ~ 0.055 | in operation |
| CERN Linac 3 | IH | $^{208}Pb^{25+}$, $^{208}Pb^{27+}$ | 101.28, 202.56 | 0.1 | 0.023 ~ 0.094 | in operation |
| CERN REX-ISOLDE | IH | radioactive ions $A/q \leq 4.5$ | 101.28 | 0.1 | 0.025 ~ 0.050 | in operation |
| HICAT | IH | $^{12}C^{4+}$, $H^+$ | 216.816 | >0.12, 1.2 | 0.029 ~ 0.121 | in operation |
| BNL New EBIS | IH | $A/q \leq 6.25$ | 100.625 | 1.7 ~ 2.0 | 0.025 ~ 0.065 | in operation |
| FRANZ | IH | $H^+$ | 175 | 50 ~ 150 | 0.039 ~ 0.063 | Waiting for operation |
| GSI p-Linac | CH | $H^+$ | 325 | 70 | 0.080 ~ 0.361 | Under construction |

For very high intensities, a high accelerating gradient is favorable to relax serious space charge effects, and it can be achieved with SC accelerating structures much more easily. Taking advantage of the feature of the long lens-free sections allowed by KONUS, the superconducting CH (Crossbar H-Type) structure overcomes the lack of multi-cell, SC accelerating structures in the low and medium $\beta$ region [10].

This study is dedicated to investigate a solution using a combination of the KONUS dynamics with the young SC CH structure for accelerating very high intensity beams in the low and medium $\beta$ range. In this solution, most of the beam acceleration will be done by the SC CH structure mainly working at $\varphi_s = 0°$ and if necessary including a short rebunching section working at negative synchronous phases typically $\varphi_s = -35°$ for improving the phase stability. The accumulated transverse RF defocusing effects can be compensated by introducing external solenoid lenses. The proposed solution is shown in FIG. 4 schematically.

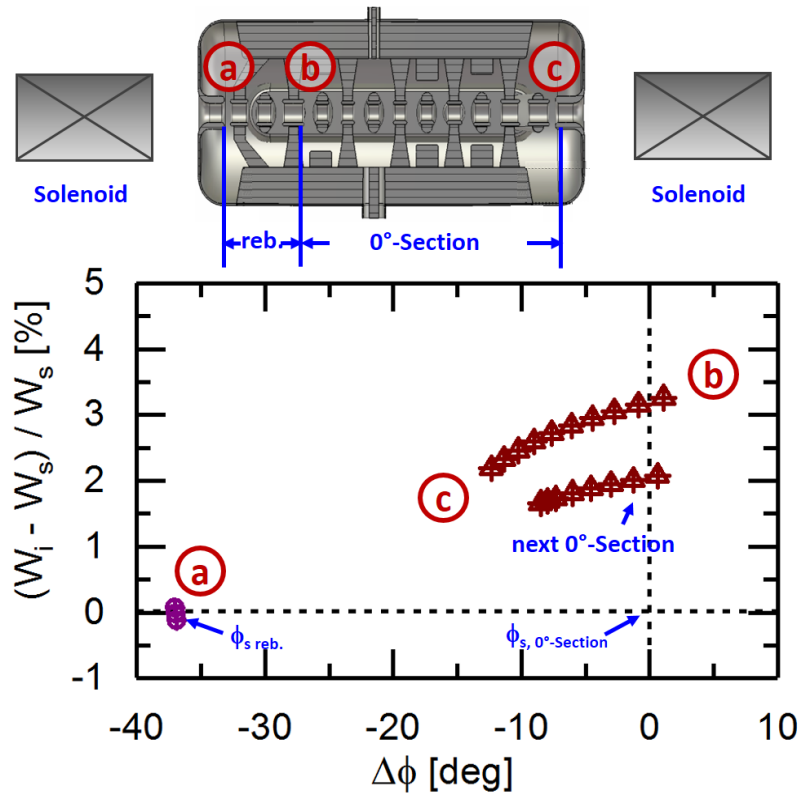

FIG. 4. A combination of KONUS and SC CH for accelerating very high intensity beams.

## II. KONUS AND SC CH DESIGN FOR A 6MW DRIFT-TUBE LINAC

To challenge very high intensity and very high power beams, a 175MHz, 150mA, CW deuteron drift tube linac aiming to increase $\beta$ from 0.07 to 0.20 is chosen for this study. The total beam power of this DTL will be considerably high as 6MW. For a convenient description, this linac will be called as the 6MW-DTL in the following text.

To avoid the safety and reliability problems e.g. activation and SC quenching which can possibly be caused by beam losses for such a very high power linac, good beam quality has to be ensured during acceleration. Therefore, a very careful design with special optimization concepts will be required for the 6MW-DTL. The primary guideline for our design is to be very conservative and to be fault-tolerant.

The considerations for the general layout of this 6MW-DTL are the follows:

- The main beam acceleration ($\beta$ = 0.10 ~ 0.20) will be accomplished by superconducting CH cavities.
- To be fault-tolerant, in front of the SC part, it's decided to add a warm transition section consisting of a 2-cell rebuncher and an IH cavity ($\beta$ = 0.07 ~ 0.10) to filter possible not-well-accelerated particles from upstream.
- Quadrupole lenses and solenoids will be used as transverse focusing elements in the RT and SC parts, respectively.

Many choices for the key parameters of its main components are based on the experience learnt from built machines and real experiments.

For the warm part, the accelerating gradient $E_a$ of the IH cavity has been taken as 2.2MV/m which is similar to another RT CW machine i.e. the FRANZ IH [11]. This is a very moderate choice for the CW operation, compared with the already reached $E_a$ values for pulsed RT IH cavities e.g. $E_a$ = 3.8, 6.4, and 4.9 MV/m for the IH-1, IH-2, and IH-3 cavities of the CERN Linac3, respectively

[12]. Concerning the quadrupole lenses, pole tip fields up to $B_{max}$ = 1.3T are available with conventional technology (room temperature, laminated cobalt steel alloys). For this linac, quadrupole lenses with $B_{max}$ = 1.15T have been adopted with a safety margin.

For the cold part, the key parameters of the SC CW linac demonstrator consisting of a 15-cell SC CH cavity and two SC solenoids can be taken as a good reference. In 2017, this demonstrator cavity accelerated up to 1.5pmA $Ar^{11+}$ beams to the design beam energy with full transmission [13]. That's the first time for CH cavities to be tested with beams. Though the demonstrator is not based on KONUS but another kind of special beam dynamics concept so-called EQUUS (EQUidistant mUltigap Structure) [14], the hardware performance proven by its experiments can be applied for KONUS lattices as well.

FIG. 5 shows the RF test results of the demonstrator cavity at 4.2K in vertical orientation without Helium vessel and in horizontal orientation with Helium vessel, respectively [15]. It can be seen that although the design accelerating gradient of this demonstrator cavity is 5.5MV/m, a maximum $E_a$ = 9.6MV/m at $Q_0$ = 8.14×$10^8$ can be achieved. Also, the two solenoids with $B_{max}$ = 9.3T have been successfully tested at 4.2K [15]. To be on a safer side, a more conservative $E_a$ = ~ 5MV/m and a relatively moderate $B_{max}$ = ~ 7T have been chosen for the SC CH cavities and the SC solenoids of the 6MW-DTL, respectively.

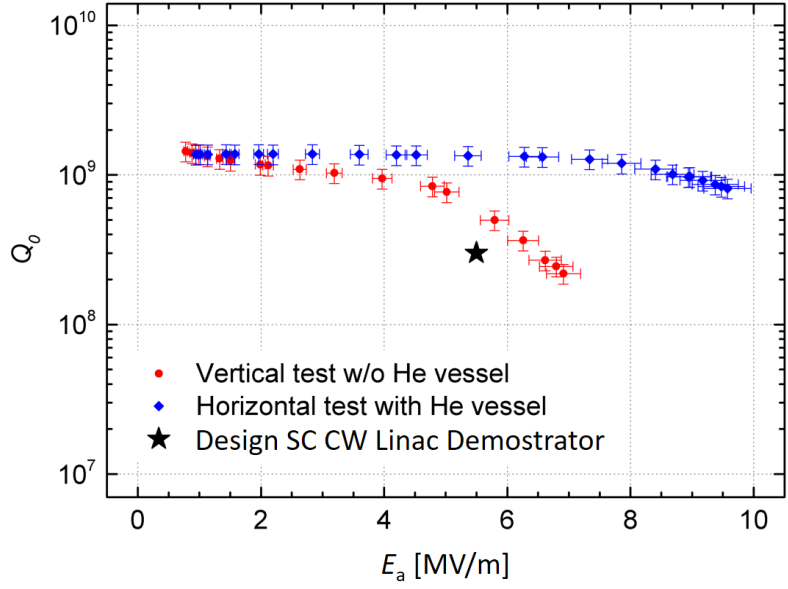

FIG. 5. RF test results of the SC CW linac demonstrator at 4.2K [15].

For easy construction and mechanical robustness, the SC CH cavities of the 6MW-DTL have been limited to be shorter than 1m, so they have relatively lower numbers of cells: 9 cells per cavity for the first two SC CH cavities and 6 cells per cavity for the others, respectively. To still take advantage of the efficiency of the 0° sections, the so-called "super 0° sections" across two cavities have been introduced. In FIG. 6, one can see that the cold part of the designed 6MW-DTL has totally five 0° sections among which the last three are the super ones. A summary of the detailed design parameters for the ~12m-long 6MW-DTL is given in TABLE II.

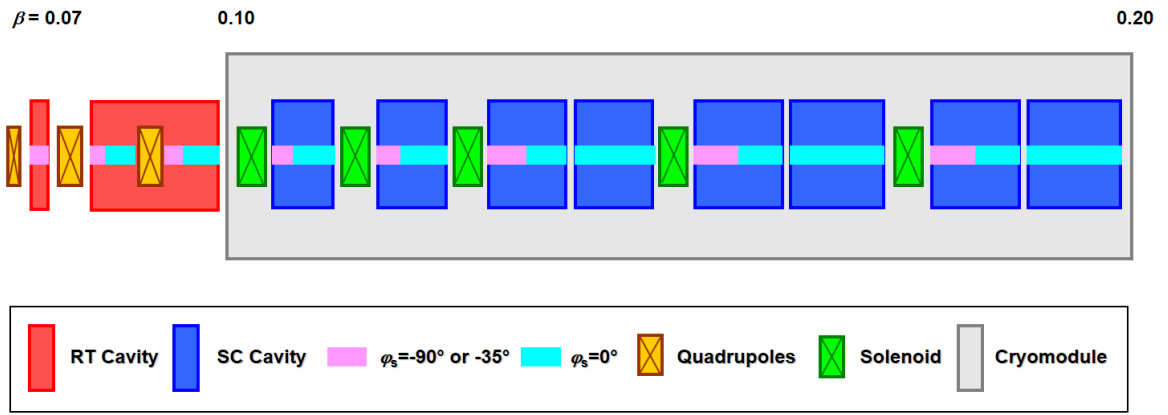

FIG. 6. Schematic layout of the 6MW-DTL.

TABLE II. Main design parameters for the 6MW-DTL.

| Parameter | Value |
|---|---|
| Frequency [MHz] | 175 |
| Beam intensity [mA] | 150 |
| Beam velocity $\beta$ | 0.07 ~ 0.20 |
| Total beam power [MW] | 6 |
| Number of RT cavities | 2 |
| Number of SC cavities | 8 |
| $E_a$ for RT IH [MV/m] | 2.2 |
| $E_a$ for SC CHs [MV/m] | ~5.0 |
| Number of accelerating cells | 76 |
| Number of multiplet lenses | 3 |
| Number of solenoids | 6 |
| $B_{max}$ for quadrupoles [T] | 1.15 |
| $B_{max}$ for solenoids [T] | 7 |
| Total layout length $L_{total}$ [m] | 12.2 |

The starting conditions $\Delta\phi$ and $\Delta W$ of the bunch at the first gap of each 0° section have decisive influence on the KONUS beam dynamics. FIG. 7 shows the typical working area used by KONUS designs: $\Delta\phi$ = -5° ~ 5° and $\Delta W$ = 3% ~ 9%. Generally speaking, a more positive $\Delta\phi$ is for a relatively longer 0° section, while a more negative $\Delta\phi$ is for a shorter 0° section; and a smaller $\Delta W$ is for relatively lower beam energy, while a larger $\Delta W$ is for higher beam energy. For the 6MW-DTL, the design choice is also quite conservative: $\Delta\phi$ around -5° and $\Delta W$ around 5%.

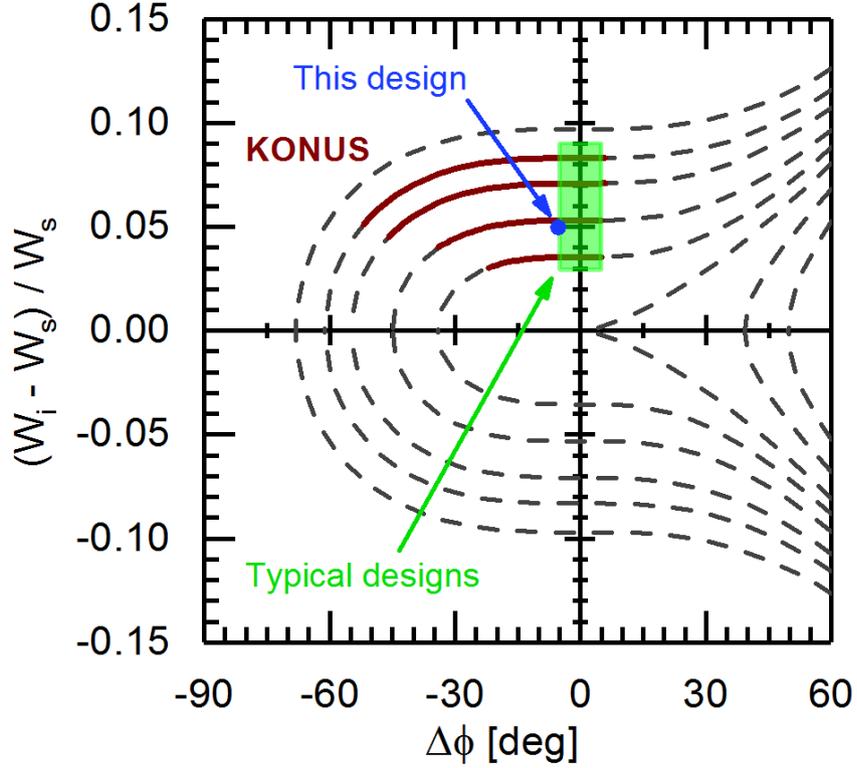

FIG. 7. Typical starting $\Delta\phi$ and $\Delta W$ for the 0° sections and the choice for the 6MW-DTL.

Last but not least, special concepts have been introduced to the design for avoiding beam instabilities induced by emittance transfer during the beam acceleration. For intense beams, the coupling-related resonances play a remarkable role in causing beam instabilities. At different emittance ratios, the growth rates and thresholds of the coupling-related resonances can be identified and visualized in the so-called "Hofmann Instability Charts" [16] by two dimensionless parameters namely the tune ratio $\sigma_l/\sigma_t$ and the tune depression $\sigma/\sigma_0$, respectively. On the charts, the shaded areas indicate the positions for emittance transfer and the developing speed of these parametric resonances. Our applied concepts are: 1) to put the starting $\sigma_l/\sigma_t$ close to the equipartitioning line; 2) to keep similar evolutions of the longitudinal and transverse phase advances along the 6MW-DTL. The controls of the phase advances are done by properly tuning the transverse and longitudinal focusing strengths. The goal is to localize the beam footprints of the 6MW-DTL in the clean i.e. safe space on a fixed Hofmann Chart with minimum number of times through the resonance peaks so that the emittance ratio between the longitudinal and transverse planes can be held almost constant.

## III. BEAM DYNAMICS SIMULATION RESULTS

The beam transport simulation along the 6MW-DTL has been performed with LORASR [17], a dedicated computer code for the KONUS dynamics using H-type structures. The input distribution including ~1 million macro-particles has been obtained from the simulation of a $\beta$=0.07 deuteron RFQ accelerator [18] at 150mA. The phase spread is ±30° and larger than the ideal range i.e. ±15° for the KONUS dynamics. FIG. 8 shows also the output phase spaces of the 6MW-DTL. It can be seen that after the acceleration, the particle distributions are still very concentrated in both transverse planes. Although some small halos can be found in the longitudinal plane, they are still in the acceptable range.

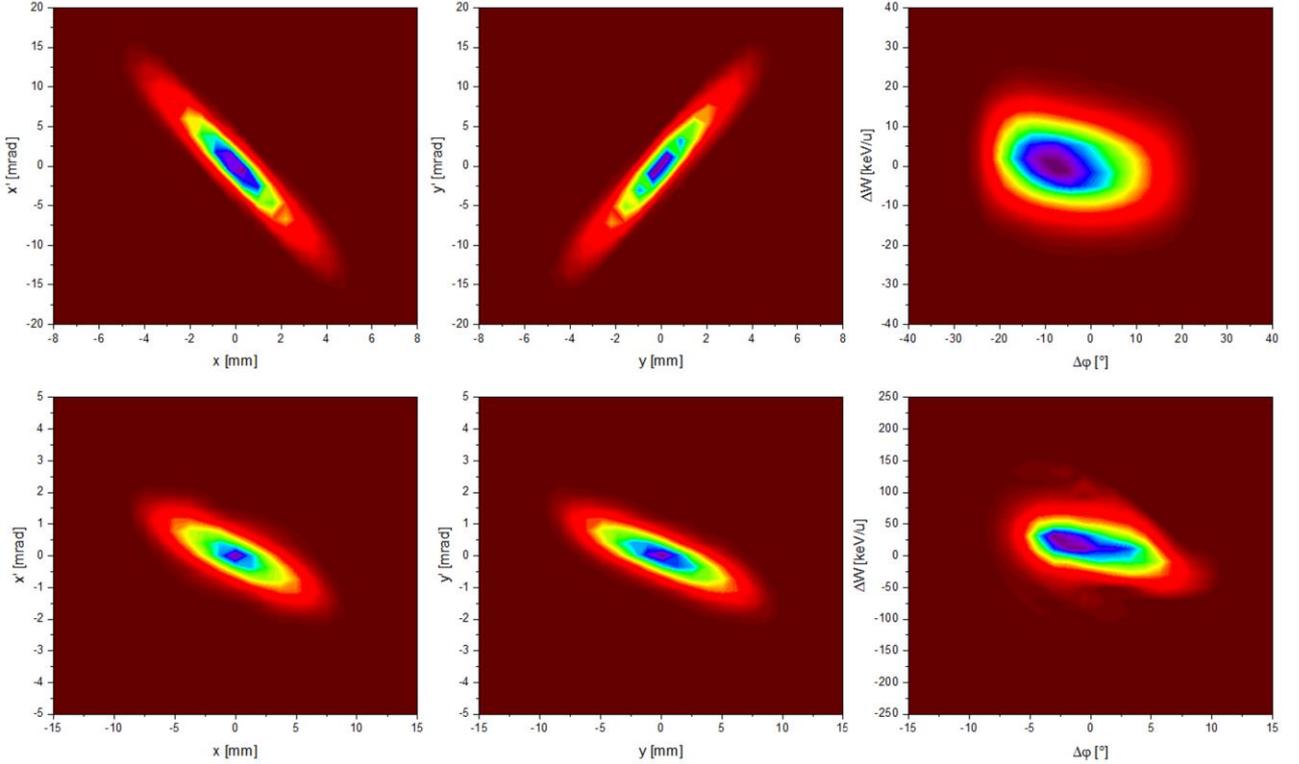

FIG. 8. Input and output particle distributions.

The motions of the beam bunch centers in the seven normal and super 0° sections are shown in FIG. 9, where the numbers are indicating the cell IDs. Around the design working point, the starting $\Delta\phi$ and $\Delta W$ values of the 0° sections are ranging from -4° to -6° and from 4.5% to 5.8%, respectively. It can be seen that all the 0° sections have been ended properly at the safe positions in the longitudinal phase space.

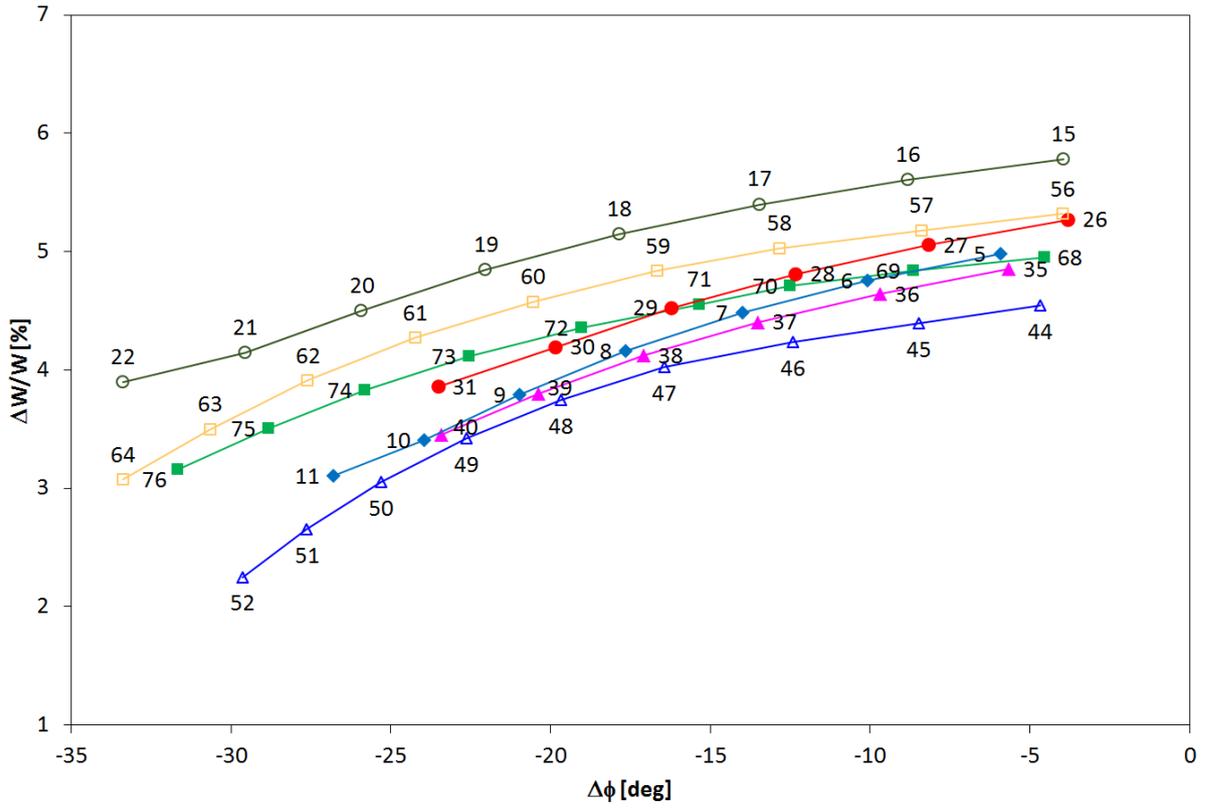

FIG. 9. Beam bunch centers (marked with the cell IDs) at all 0 °cells.

The transverse and longitudinal beam envelopes are plotted in FIG. 10 and FIG. 11, respectively. In the transverse planes, the evolutions of the beam sizes are almost identical, especially in the cold part. Besides, the relative energy spread and the phase spread of the beam are also well confined throughout the DTL.

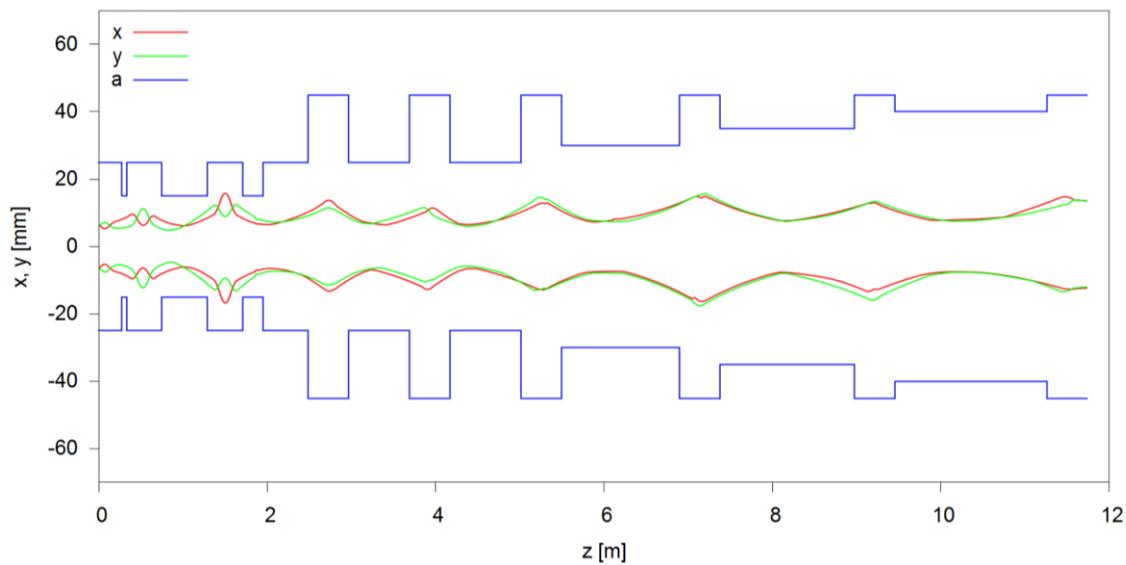

FIG. 10. Transverse beam envelopes.

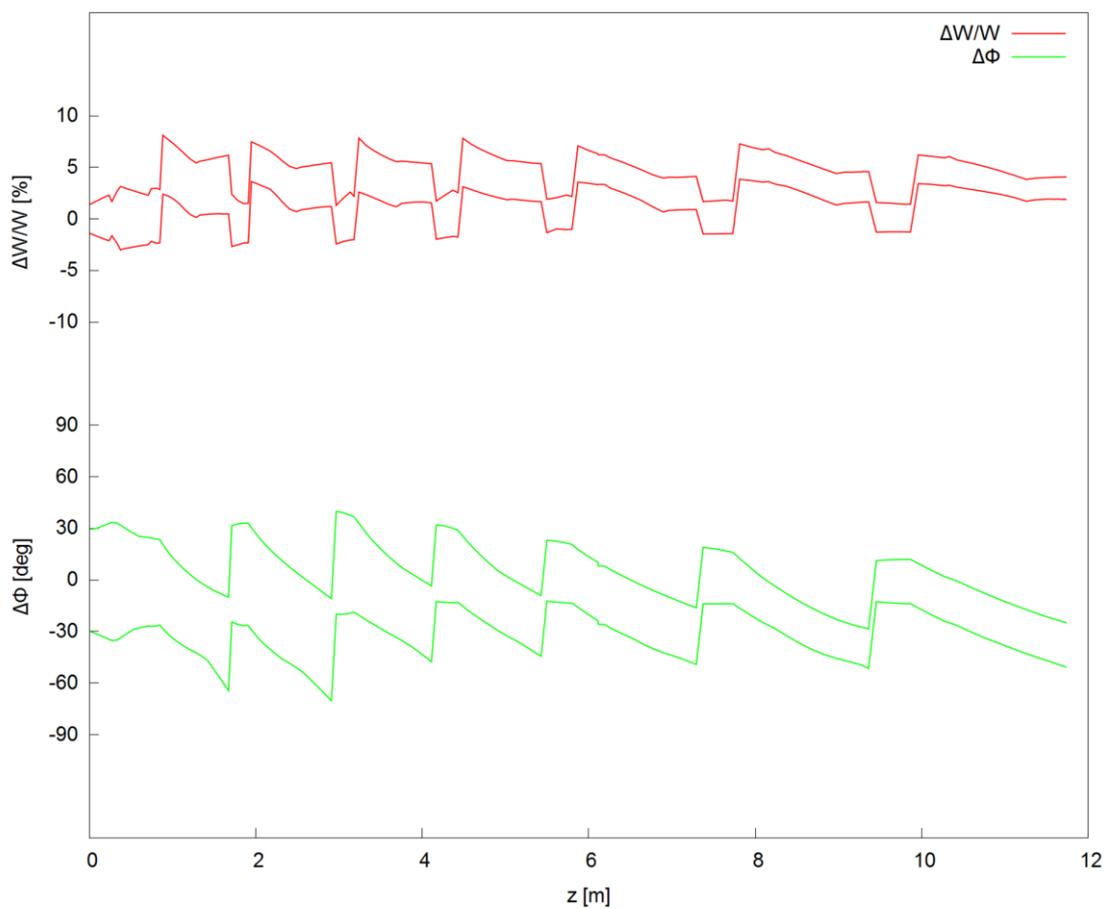

FIG. 11. Longitudinal beam envelopes.

FIG. 12 shows the ratios of the longitudinal and transverse emittances and phase advances along the accelerating channel, where the numbers indicate the IDs of the lattice periods. In LORASR, each lattice period consists of an arbitrary number of accelerating cells with no focusing elements in between, followed by an arbitrary number of focusing elements (quadrupole lenses or solenoids). Each period ends behind the last lens of the corresponding array. It can be seen that the tune ratio is nearly constant along the accelerating channel, except one small jump in the second lattice period i.e. between the first two lenses. This is because the given input phase spread is twice as big as the ideal one for KONUS and a strong 2-cell rebuncher working at $\varphi_s = -90°$ is necessary there for longitudinal matching. Afterwards, this situation has been changed back to normal by the properly designed second lens. As expected, the emittance ratio has been held relatively stable at ~1.75.

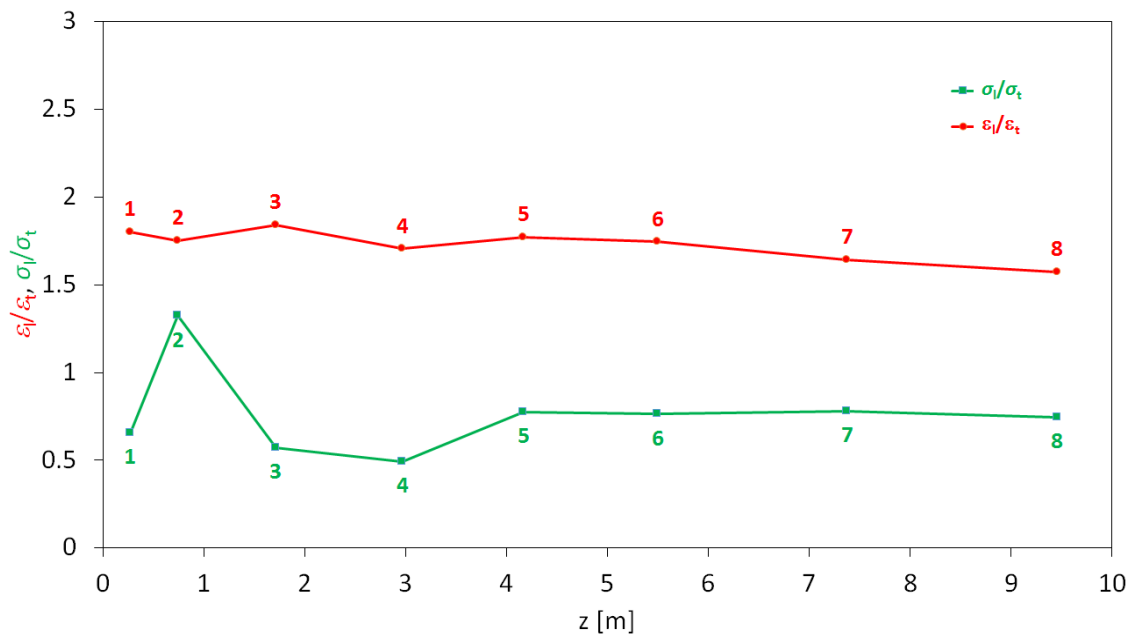

FIG. 12. Ratios of emittances and phase advances (marked with the IDs of the lattice periods).

FIG. 13 shows the beam footprints of the 6MW-DTL on the $\frac{\varepsilon_l}{\varepsilon_t} = 1.75$ Hofmann Chart. The Hofmann Chart itself has been generated using the TraceWin code [19]. It can be seen that most of the beam trajectories are well confined in the clean area around the equipartitioning line on the chart, except a short trip into the neighboring resonance peak in the second lattice period (this is caused by the rebuncher as mentioned in the previous paragraph). In this way, good beam quality has been ensured.

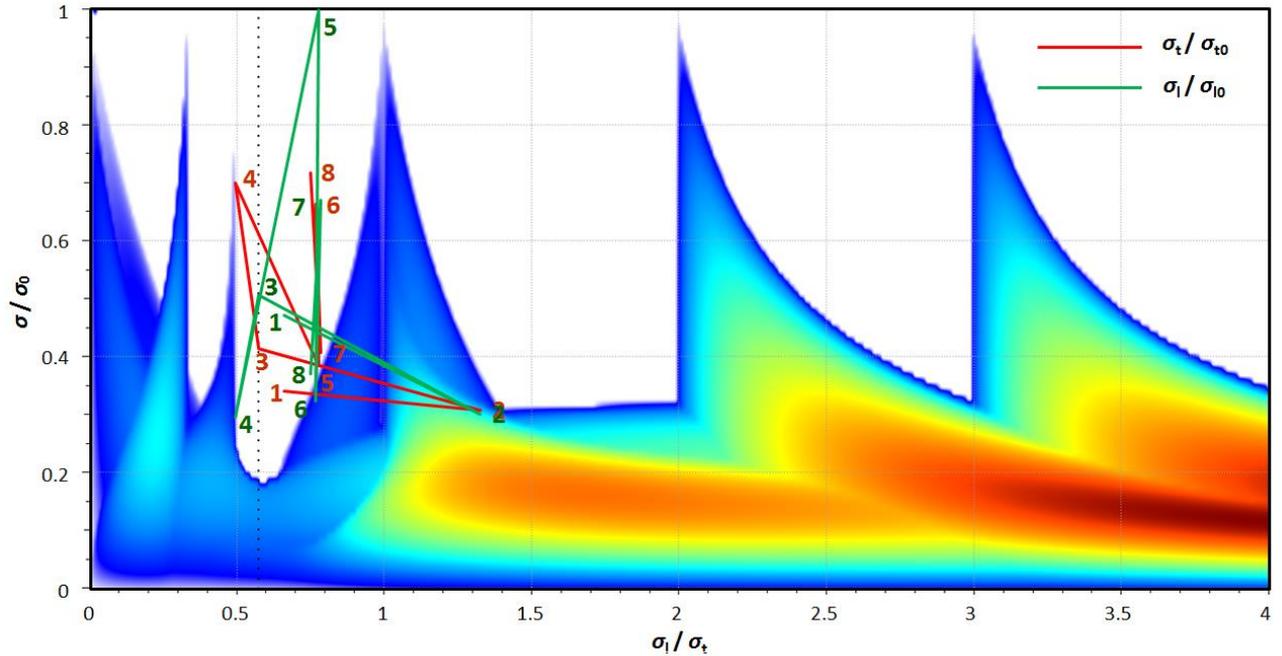

FIG. 13. Beam footprints of the 6MW-DTL on the $\frac{\varepsilon_l}{\varepsilon_t}$ = 1.75 Hofmann Chart.

## IV. ERROR STUDIES

For the design stage, only perfect accelerator components and ideal operating conditions have been taken into account. In reality, however, more or less perturbations to the design case are inevitable. As required, the design should be fault-tolerant. Therefore, systematic studies have been done to check the performance of the 6MW-DTL in the presence of various possible errors.

For the LORASR code, the following kinds of errors can be implemented [20, 21]:

- Transverse offsets of magnetic lenses (LOFF).
- Rotations of magnetic lenses in all directions (LROT).
- Voltage amplitude errors for accelerating cells and tanks (VERR).
- Phase errors for tanks (PERR).

TABLE III shows the three error settings applied to the 6MW-DTL design. From Setting-1 to Setting-3, the maximum errors have been increased accordingly.

TABLE III. Maximum errors used for the three settings.

| Error Type | Setting-1 | Setting-2 | Setting-3 |
|---|---|---|---|
| LOFF [mm] | 0.1 | 0.2 | 0.3 |
| LROT [mrad] | 1.0 | 2.0 | 3.0 |
| VERR [%] | 1.0 | 2.0 | 3.0 |
| PERR [°] | 1.0 | 2.0 | 3.0 |

To model various non-ideal cases in the simulation, the above-mentioned errors have been generated randomly and mixed with each other. In LORASR, the randomly-generated errors are Gaussian distributed and truncated at the maximum $A=\pm 2\sigma_d$ with $\sigma_d$ being the standard deviation. FIG. 14 shows the alignment errors of the lenses (each singlet has been treated separately) applied for Batch-1, Batch-2, and Batch-3 (marked in red, green, and blue, respectively), as an example. For each error setting, a batch of 100 runs for different cases has been performed using ~1 million macro-particles.

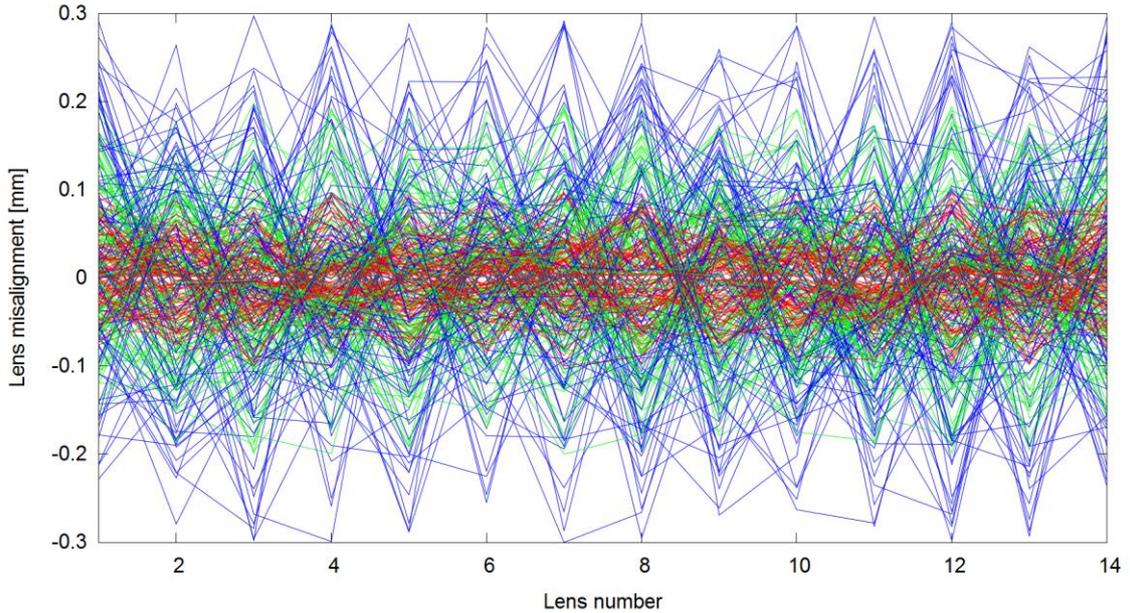

FIG. 14. Alignment errors of the singlets and solenoids.

The transverse and longitudinal beam envelopes of all 300 non-design DTLs are shown in FIG. 15, where the nominal case is marked in yellow for comparison. It is clear that the larger the errors, the worse the situation. For Batch-1, the transverse and longitudinal beam sizes have very limit deviations to the reference case. For Batch-2, though very few runs have produced some particles starting to leave the bunch longitudinally, no particle touches the drift tubes or lenses, so there are still no beam losses.

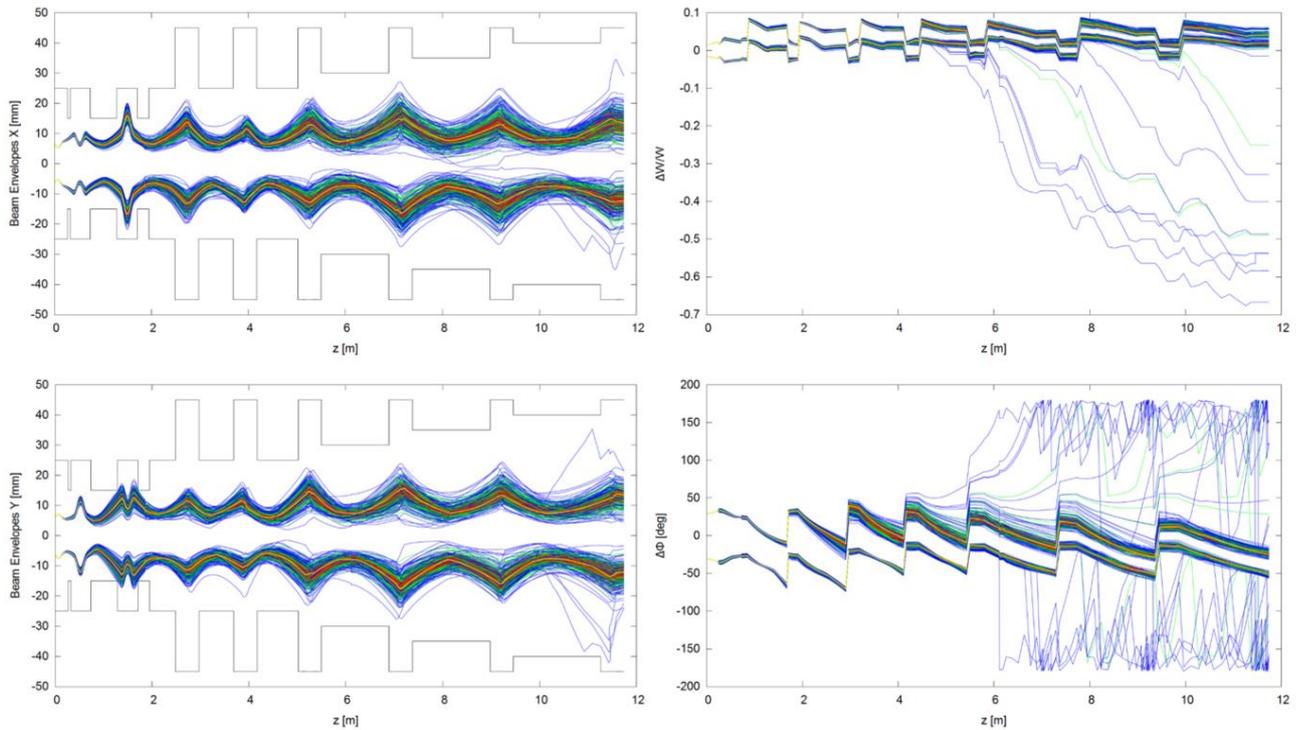

FIG. 15. Beam envelopes with errors (red: Batch-1, green: Batch-2, blue: Batch-3).

For Batch-3, some beam losses have been observed. More clearly one can see from FIG. 16 that only a few runs have beam losses and the lowest transmission efficiency is still better than 99.997%. Furthermore, most of the lost particles happened in the RT part. Therefore, the 6MW-DTL is still robust enough against errors.

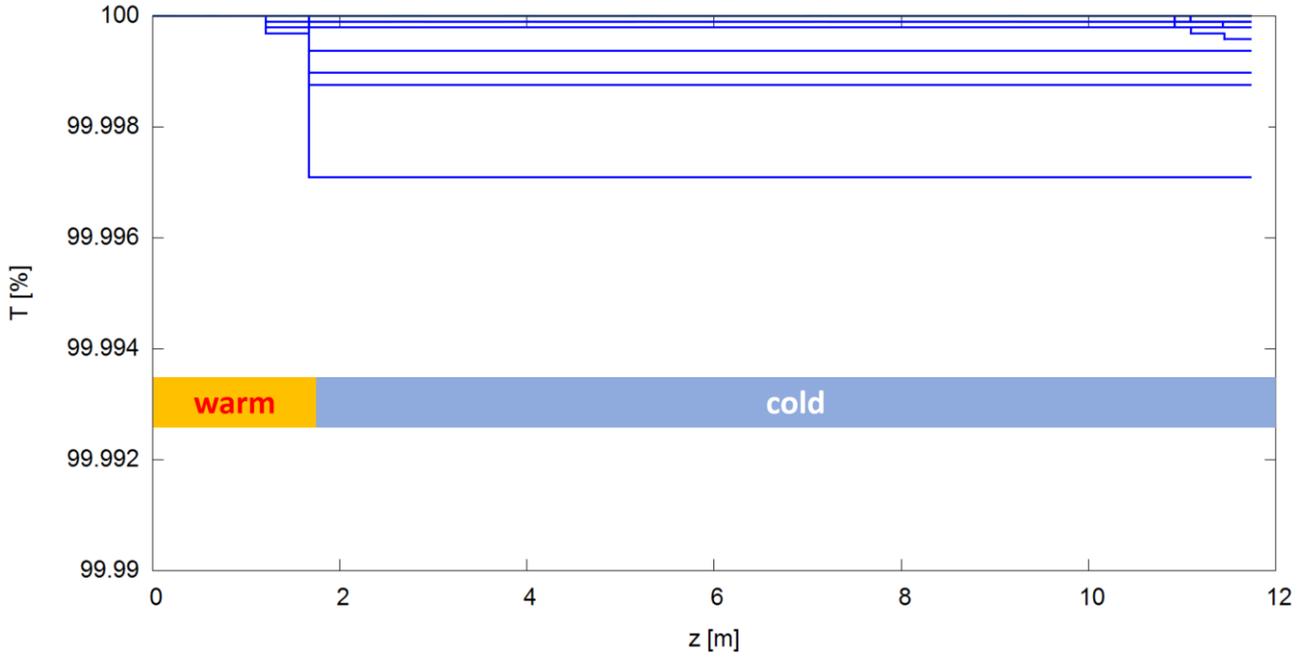

FIG. 16. Transmission efficiencies with errors (red: Batch-1, green: Batch-2, blue: Batch-3).

FIG. 17 shows the statistics of the additional emittance growths induced by the mixed errors. The definition for the additional emittance growth is given as below:

$$\Delta\varepsilon_{addi.} = \frac{\varepsilon_{out}^{with\ errors} - \varepsilon_{out}^{without\ error}}{\varepsilon_{in}} \qquad (2)$$

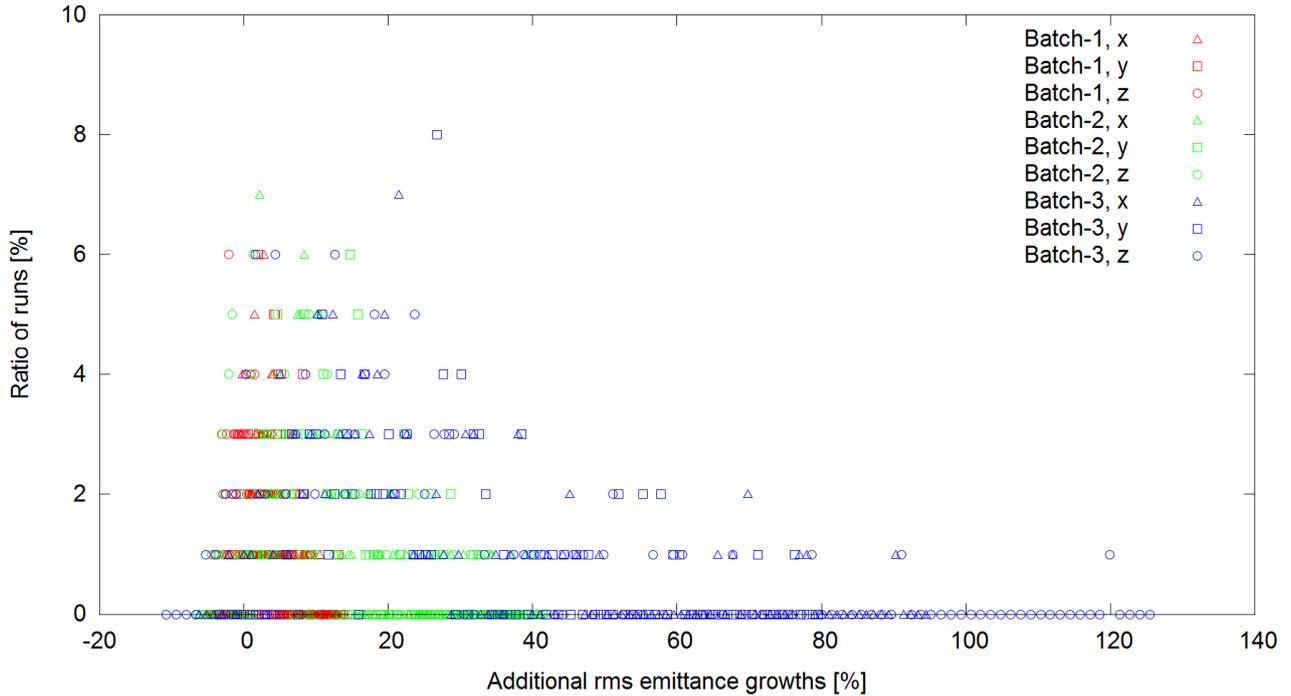

FIG. 17. Additional emittance growths induced by the errors.

For Batch-1 and Batch-2, the maximum additional emittance growth is <45%, while for Batch-3, $\Delta\varepsilon_{addi.}$ is <95% and <130% in the transverse and longitudinal planes, respectively. A further particle tracing study proves that those large longitudinal additional growths are contributed by only a few off-energy particles. Generally speaking, therefore, the beam quality stays still good in the presence of mixed errors.

## V. CONCLUSIONS

In the last several decades, the combination of KONUS dynamics strategy and room temperature H-type structures has been developed as an efficient solution for accelerating low and medium $\beta$ beams. To challenge very high intensity beams, a new idea to combine the KONUS dynamics with the young SC CH structure has been proposed and investigated.

The investigation is based on a 150mA, 6MW deuteron linac. To ensure a safe and reliable CW operation at such a very high intensity and mainly at 4.2K, 1) a carefully design guideline "to be very conservative and to be fault-tolerant" has been followed; 2) very conservative design choices e.g. $E_a$ and starting $\Delta\phi$ and $\Delta W$ for the 0° sections have been adopted; 3) and special design concepts e.g. adding a warm transition section, introducing "super 0° sections", and trying to confine the beam footprints inside the largest safe area on the Hofmann Chart have been applied.

Due to the feature of the long lens-free sections allowed by the KONUS dynamics and the high accelerating gradient provided by the SC CH structure, a very compact layout (only ~12m long) with a low number of accelerator components has been realized for the 150mA, 6MW DTL by the new solution. Detailed analyses show that the emittance transfer has been minimized and good beam quality has been ensured. In addition, benefitting from the fewer components, the design shows also large tolerances against possible errors.